\documentclass[twocolumn]{aastex631}

\usepackage{graphicx}
\usepackage{comment}
\usepackage{isotope}
\usepackage{subfigure}

\newcommand{\Msun}{M$_\odot$}
\newcommand{\rsun}{R$_\odot$}
\newcommand{\teff}{$T_\mathrm{eff}$}
\newcommand{\teffa}{$T_\mathrm{eff,A}$}
\newcommand{\teffb}{$T_\mathrm{eff,B}$}
\newcommand{\loglum}{$\log\left(\frac{L}{L_{\odot}}\right)$}
\newcommand{\logluma}{$\log\left(\frac{L_A}{L_{\odot}}\right)$}
\newcommand{\loglumb}{$\log\left(\frac{L_B}{L_{\odot}}\right)$}

\newcommand{\MA}{$M_\mathrm{A}$}
\newcommand{\MB}{$M_\mathrm{B}$}

\begin{document}

\title{Analysis of DQZ White Dwarf Evolution through Procyon}

\author[0000-0002-0786-7307]{Momin Y. Khan}
\affiliation{Department of Physics and Astronomy, Baylor University, Waco, TX 76798-7316} 
\email{momin\_khan1@baylor.edu}

\author[0000-0001-7010-7637]{Barbara G. Castanheira}
\affiliation{Department of Physics and Astronomy, Baylor University, Waco, TX 76798-7316}


\shorttitle{Modeling Procyon}
\shortauthors{Khan \& Castanheira}

\begin{abstract}
Procyon is a great system to probe stellar evolution of non-interacting binaries.  We present an extensive grid of MESA (Modules for Experiments in Stellar Astrophysics) evolutionary tracks to constrain the evolution of Procyon~A and B. We systematically vary the initial parameters of our grid anchored by precise dynamical masses and spectroscopic determinations of effective temperature (\teff) and luminosity ($L$) to match the stars' positions on the H-R Diagram. 
Our goal is two-fold: (i) to quantify how the inferred system age and the progenitor mass of Procyon~B depend on metallicity ($Z$), mixing length ($\alpha$), and core overshoot ($\beta$), and (ii) to determine the best fitting model of Procyon~B within a model-based initial-to-final mass relationship (IMFR) for hydrogen-deficient white dwarfs. 
Our best-fit models reproduce the observed properties for both components, yielding $M_\mathrm{A}=1.487 \pm 0.095$\,\Msun, $M_\mathrm{B}=0.592\pm 0.082$\,\Msun, a system age of $2.23 \pm 0.90$ Gyr, and a white dwarf cooling age of $1.20\pm0.49$\,Gyr for Procyon~B, consistent with independent determinations. 
Our results point to higher core overshoot than the standard adopted range, with the best fits ranging from $\beta=0.5-1.0$. From our model grid, we map Procyon B to the initial-to-final mass relationship for H-deficient white dwarfs in the $1.9\!-\!2.6$\,\Msun\ progenitor range. Additionally, we implement the accretion of heavy metals onto the surface of the WD and fit our isotopic abundances to spectroscopic observations. We outline the physics used in our analysis.



\end{abstract}


\section{Introduction} \label{intro}
Procyon is one of the most notable and well-studied binary systems, comprised of $\alpha$ Canis Minoris, a main sequence F5 IV-V star, and white dwarf~0736+053, a white dwarf companion with spectral type DQZ (carbon and metal lines), Procyon~A and B, respectively. 
\cite{1844MNRAS...6R.136B} observed that the proper motion of Procyon in declination was not constant, which was later confirmed to be caused by an unseen companion \citep{1896PASP....8..314S}. Members in a large orbital period and separation like Procyon~A and B are classified as a ``Sirius-like system'', wide binaries containing a main sequence star and a white dwarf. In these systems, stars are likely to evolve without mass transfer, making them incredibly important objects in probing stellar evolution models, since they allow us to place independent constraints to the initial-to-final-mass relation (IFMR). Furthermore, adding the white dwarf cooling time to the main-sequence lifetime yields the total age of the system. 

Procyon~A is the eighth brightest star in the night sky, with a visual magnitude of $0.34$ \citep{1953ApJ...117..313J}. Meanwhile, its dim companion has a measured magnitude of $V=11.3\pm0.1$ \citep{1994PASP..106..356W} or a computed magnitude $V=10.82 \pm 0.03$ 
\citep{2002ApJ...571..512H}. The orbital solution has a period of $40.840 \pm 0.22$\,years, with an eccentricity of $0.39785 \pm 0.00025$, and a semi-major axis of $4.3075 \pm 0.0016 $\,arcsec \citep{2015ApJ...813..106B}. The two stars are never apart by more than 5'', making the faint Procyon~B historically difficult to image directly \citep{2015ApJ...813..106B}. 

In ``Sirius-like systems'' such as Procyon, dynamical masses and precise photospheric parameters provide strong anchors for stellar models. Despite extensive previous studies of Procyon, there remains value in a homogeneous, system-level benchmark that (i) jointly fits both components with a single set of assumptions; (ii) maps the sensitivity of inferred ages and progenitor masses to metallicity ($Z$), mixing length ($\alpha$), and core overshoot ($\beta$); and (iii) clarifies which pieces of physics are essential to model a hydrogen-deficient white dwarf.

In this paper, we present our modeling and analysis of the Procyon binary system, using self-consistent evolutionary models. We follow the systematic study done by \cite{2024ApJ...977...41K} for Sirius. We discuss the choices in the parameter space of the model grid calculated for the stellar evolution of Procyon~A and B and present our best  model. Our goals are to quantify how $Z$, $\alpha$, and $\beta$ translate into uncertainties in age and final mass, and to place Procyon~B onto an IFMR tailored to hydrogen-deficient white dwarfs.
We then compared our results with previous studies, establishing external uncertainties in our determinations. Our goal is not only to provide a better understanding of the evolution of Procyon~A and B but also to shed light on modeling the evolution of binary systems as a whole.

\section{Observational Properties of Procyon} \label{observ}

In this section, we briefly review the relevant observations of Procyon, highlighting the astrophysical quantities derived in previous studies that we used to define the boundaries of our computed model grid. 

\subsection{Astrometric measurements} \label{astrometric}

Astrometric solutions rely on the precise positions and motions of celestial bodies and their parallax. Unfortunately, there is no Gaia parallax determination for Procyon~A as it is brighter than $G_\mathrm{Gaia}=3$, nor for B, because of the small angular separation between the two (less than 5'') and because it is $\sim 10$\,magnitudes fainter than A. The weighted mean absolute parallax for Procyon~A is $0.2850\pm0.0007$\,arcsec, obtained from the compiled historical measurements of $0.2832 \pm 0.0015$\,arcsec from USNO plates \citep{Girard_2000}, $0.2846 \pm 0.0013$\,arcsec from Hipparcos \citep{2007A&A...474..653V}, and $0.2860 \pm 0.0010$\,arcsec from Multichannel Astrometric Photometer \citep{2006AJ....131.1015G}. These determinations yield a distance of $3.509\pm0.009$\,pc from Earth. 

Mass is the most important quantity when it comes to constraining stellar evolution. In binary systems, we can determine the most precise mass values of the components from the orbital solution. We have identified three recent independent astrometric studies of Procyon. \cite{Girard_2000} combined 83\,years of data obtained with photographic plates and with the Planetary Camera of the Hubble Space Telescope (HST) to derive masses of \MA$=1.497\pm0.037$\,\Msun\ and \MB$=0.602\pm0.015$\,\Msun\,, where \MA\ and \MB\ are the masses of Procyon~A and B, respectively. From the analysis of observations with the Multichannel Astrometric Photometer between 1986 -- 2004, \cite{2006AJ....131.1015G} derived \MA$=1.43\pm0.034$\,\Msun\ and \MB$=0.58\pm0.014$\,\Msun. The newest determinations used images obtained over two decades with the HST and historical measurements dating back to the 19th century to derive \MA$=1.478\pm0.012$\,\Msun\ and \MB$=0.592\pm0.006$\,\Msun\ \citep{2015ApJ...813..106B}.

There are many reasons to assume the stars evolved without significant mass transfer: the primary ones being separation, eccentricity, and metallicity. The current separation of the system is very wide, with a periastron of $\sim9$\,AU \citep{2015ApJ...813..106B}. Additionally, the eccentricity is very high, meaning that it is unlikely that there was a point in the binary's lifetime where they developed a common envelope, as this would circularize the orbits \citep{2015ApJ...813..106B}. Beyond that, there is no spectroscopic evidence of pollution in Procyon~A's photosphere \citep{1982A&A...113..135K}, which would indicate mass transfer during the ejection of the planetary nebula of B. 

Under this assumption and using the upper limit for stellar wind of $2\times10^{-11}$\,\Msun\ yr$^{-1}$ from radio observations \citep{1993ApJ...406..247D}, it is safe to assume that the mass of Procyon~A has not change significantly throughout its evolution. Therefore, we limited the calculations of our model grid for its mass for values between 1.4 and 1.53\,\Msun, based on astrometry measurements. 

The observed mass for Procyon~B is within the range 0.566 -- 0.617\,\Msun, which represents its final mass. We used the semi-empirical IFMR for hydrogen-deficient white dwarfs \citep{Barnett_2021} to constrain the initial mass of the progenitor of B in our model grid to values between 1.9 and 2.6\,\Msun.

\subsection{Photospheric Parameters} \label{photospheric}

The effective temperature (\teff) and luminosity ($L$) of Procyon~A and B are additional observational physical quantities that allow us to constrain the space parameters of our grid of evolutionary models. 

\subsubsection{Photospheric observations of Procyon~A}

\cite{2005ApJ...633..424A} analyzed the spectrophotometric data taken from the International Ultraviolet Explorer and the HST Space Telescope Imaging Spectrograph, along with broadband photometry deriving \teffa$=6543 \pm 84$\,K, and \logluma$=0.83 \pm 0.04$ for Procyon~A. In a subsequent work, \cite{2012A&A...540A...5C} created synthetic stellar surfaces to re-analyze the data from \cite{2005ApJ...633..424A} for different wavelengths and with updated opacities, deriving \teffa = $6591 \pm 43$\,K or \teffa = $6556 \pm 84$\,K, depending on the source of bolometric flux used. This resulted in a slightly higher luminosity of \logluma $=0.843$. An independent study by \cite{2021AJ....162..198B}, measured the diameter of Procyon~A using the Navy Precision Optical Interferometer and combined with Hippacos parallax to determine \teffa$=6548\pm47$\,K and \logluma$=0.84\pm0.02$. 

Our brief review of published values is not complete by any means, but indicates that independent methods yield consistent values, within the error bars. Therefore, it is straightforward to choose \teffa=$6550$\,K and \logluma$=0.84$ as target values for effective temperature and luminosity when modeling the evolution of Procyon~A. 

\subsubsection{Photospheric observations of Procyon~B}

One of the first complete studies on Procyon~B was done by \cite{1997ApJ...480..777P}, where they obtain U, B, V, R, and I magnitudes with the HST Wide Field and Planetary Camera 2. By fitting the observed energy distribution to photospheric models, they derived \teffb$=8688 \pm 200$\,K. In a later study however, \cite{2002ApJ...568..324P} fitted spectra from the HST Space Telescope Imaging Spectrograph, deriving \teffb $=7740 \pm 50$\,K. The reason for this significantly cooler temperature was that the spectroscopic data revealed the presence of carbon, magnesium, calcium, and iron in the photosphere, influencing the opacity of helium and the determination of \teff. \cite{2012ApJS..199...29G} determined the photospheric parameters of nearby white dwarfs, including Procyon~B, by combining the B, V, R, and I photometry \citep{1997ApJ...480..777P} and parallax to fit their optical spectra, obtaining \teffb $=7871 \pm 433$\,K. More recently, \cite{2016MNRAS.462.2295H} derived \teffb$=7876\pm433$\,K, after the inclusion of 3-D corrections to the models similar to the ones described in \cite{2013A&A...559A.104T}. 

As for the literature review of Procyon~A, our list of effective temperature values for B is not complete, but indicates consistency of the independent determinations, except for the hottest value derived from models without metals \citep{1997ApJ...480..777P}. Based on these previous studies, we computed evolutionary models for the white dwarf in the cooling sequence down to \teffb$\sim7800$\,K.  

The luminosity of Procyon~B of \loglumb$=-3.31$ is explicitly listed in \cite{2012ApJS..199...29G}, calculated from the photometric observations \citep{1997ApJ...480..777P} with models that included metal pollution \citep{2002ApJ...568..324P}. As \teff\ is consistently $\sim7800$\,K from independent determinations, and radius is model dependent, the computed values of luminosity are nearly the same. In our models, we used \loglumb$=-3.31$ as the target luminosity of Procyon~B. 

\subsection{Radius Measurements} \label{radius}

The only component of the Procyon system for which the angular size was measured directly is A. 
\cite{2005ApJ...633..424A} observed the limb darkening of Procyon~A, using K band from Very Large Telescope Interferometer (VLTI) and Mark III interferometry in the optical,  
combined with parallax information, to derive $R_{\mathrm{A}} = 2.031 \pm 0.013$\,\rsun. Using the VLTI commissioning instrument in the K band, \cite{2004A&A...413..251K} measured the limb darkened angular diameter, and with the available parallax, resulted a radius of $R_\mathrm{A}=2.047\pm0.020$\,\rsun. Measurements obtained with the Navy Precision Optical Interferometer and parallax result in $R_\mathrm{A}=2.04\pm0.01$\,\rsun\ \citep{2021AJ....162..198B}. These independent measurements are consistent with each other and provide another important observational constraint. 

The radius for Procyon~B is a derived quantity. Therefore, it is not used to limit the boundaries of our model grid. For completeness, we list a couple of values from different derivations. From the combination of \teff\ and $L$, \cite{2002ApJ...568..324P} obtained $R_\mathrm{B}=0.01234 \pm 0.00032$\rsun, which is independent of mass-radius relationship. Alternatively, from their astrometric solution, \cite{2015ApJ...813..106B} published $R_\mathrm{B} =0.01232 \pm 0.00032$\,\rsun. 

\subsection{Chemical Composition} \label{metals}
Metallicity plays an important role in the evolution of stars, influencing age, luminosity and effective temperature. Most studies of Procyon~A concluded that its metallicity is close to solar.
Using the Tull Spectrograph at the Harlan J. Smith 2.7\,m telescope at McDonald Observatory, \cite{2002ApJ...567..544A} 
derived an abundance of $\log\epsilon(\mathrm{Fe}) = 7.36 \pm0.03\,(\sigma = 0.15)$\,dex  from Fe I and Fe II lines, using one-dimensional and three-dimensional analyses, making it about 0.5\,dex more iron deficient than the Sun. In a later study, with the analysis of 22 chemical elements, \cite{2004A&A...420..183A} derived a solar-like abundance of $\log\epsilon(\mathrm{Fe}) = 7.63$. 

A comprehensive study of metallicity was conducted by \cite{2014A&A...564A.133J} for the Gaia FGK benchmark stars, which included Procyon~A. Their goal was to calibrate automatic pipelines to determine the photospheric parameters of the stars. They used up to seven different methods to analyze high resolution and high signal-to-noise spectra with different instruments: HARPS, NARVAL, and UVES. They derived a value close to solar metallicity, with a very small scatter ($[\mathrm{Fe}/\mathrm{H}]=+0.01$ with $\sigma\,\mathrm{Fe I}=0.01$ and $\sigma\,\mathrm{Fe II}=0.02$). 

Photometric observations of Procyon~B suggested a helium photosphere \citep{1997ApJ...480..777P}, which was later confirmed with spectroscopic observations  \citep{2002ApJ...568..324P}, revealing also the presence of carbon [$\log\mathrm{N(C)}/\mathrm{N(He)}=-5.5$], but also metals, such as magnesium, calcium, and iron in the photosphere. Procyon~B is therefore a rare DQZ white dwarf. Due to short timescales for diffusion, when metals are present in white dwarf photospheres, they are likely from the disruption of planetary bodies \citep{1986ApJ...302..462A}. 

Based on these previous studies, we will assume a range of metallicity for Procyon system close to solar values \citep{asplund_2009}. We calculated models with a wide range of metallicities of  $0.0125 \leq Z \leq 0.0275$ in steps of $\Delta Z=0.003$.

\section{Modeling Procyon with MESA} \label{models}

We created a grid of evolutionary models for Procyon using the Modules for Experiments in Stellar Astrophysics, or MESA \citep{2011ApJS..192....3P,2013ApJS..208....4P,2015ApJS..220...15P,2018ApJS..234...34P,2019ApJS..243...10P}. MESA is an open-source, robust, 1D stellar evolution code that utilizes modules containing equations of state, opacity, reaction rates, among other physical inputs to model stellar evolution. The {\it inlists} and MESA specifications used for our study of the Procyon system can be found at \dataset[https://zenodo.org/records/17361809]{https://zenodo.org/records/17361809}. 


We evolved stellar models starting from the zero age main sequence (ZAMS) up to the current position of the stars in the H-R diagram, based on their luminosities and effective temperatures (see Section~\ref{photospheric}). For Procyon~A, we used initial masses in the range $1.411 \leq M \leq 1.531$ \Msun\, with steps of 0.02\,\Msun. The models of Procyon~B were calculated through the thermally pulsating asymptotic giant branch, and until the white dwarf cooling phase, with initial masses in the range $1.9 \leq M \leq 2.6$ \Msun\ with steps of 0.1\,\Msun\ (see Section~\ref{astrometric}). 


\subsection{Convective Mixing \& Overshoot}

MESA uses the formalism of the mixing length theory to describe convection \citep{1958ZA.....46..108B}, where the mixing length $l$ is scaled in the convection zone to the local pressure scale height $H_\mathrm{p}$ by the free parameter $\alpha$, so that $l=\alpha H_\mathrm{p}$. $\alpha$ values can vary from star to star, depending on radii, photospheric parameters, among others, making it difficult to adopt a standard value when constructing evolutionary tracks. Overshoot is the extension of the region of mixing by a distance $\beta H_\mathrm{p}$, where $\beta$ (the overshoot) is another free parameter.  

Asteroseismology studies of Procyon~A's solar-like oscillations \citep{2008ApJ...687.1180A, 2010ApJ...713..935B} suggest that this star is in the hydrogen core-burning phase with an extensive mixing outside the convective core, due to overshoot. 
\cite{Guenther_2014} combined asteroseismic data with previous measurements of mass and photospheric parameters, determining that the best models are for a core overshoot of $1.0 \leq \beta \leq 1.5$ and mixing length of $\alpha = 1.8 \pm0.1$ for Procyon~A. We adopt a larger range of $1.7 \leq \alpha \leq 2.5$ to study the importance of effective mixing on the models.

Standard adoption for the core overshoot is low, at $\sim$ 0.2 pressure scale heights \citep[e.g.][]{Montalban_2013}. However, when fitting models of Procyon to the asteroseismic data, \cite{Guenther_2014} found that models with small overshoot did not agree well with observations. Without moderate to large overshoot, the age of many Procyon B models with $M_i \gtrsim 2.0$ is less than 2 Gyr, inconsistent with previous findings. Other studies that have modeled the system \citep[e.g.][]{2015ApJ...813..106B} have come to similar conclusions in regards to large overshoot and the system age. Therefore, we vary $\beta$ from 0.0 - 1.0 in steps of 0.5, still including models without core overshoot for completeness. 

\subsection{Wind \& Microphysics}

As discussed in Section~\ref{astrometric}, stellar wind is irrelevant in modeling Procyon~A. 
For Procyon~B, we implement Reimers' wind loss during the red giant branch \citep{1975MSRSL...8..369R} and Bloecker's during the asymptotic giant branch \citep{1995A&A...299..755B}, with scaling factors of 0.5 and 0.1 respectively. Since Procyon~B will be evolved until it reaches the white dwarf cooling phase, our models include thermal pulsations during the asymptotic giant branch, which includes both mass loss and core mass increase. 

After evolving our models past the asymptotic giant branch and removing the outer layers to form the white dwarf, we remove the hydrogen envelope of the white dwarf through a wind mass loss function and stop the run when the surface hydrogen abundance reaches log$(\frac{H}{He})=-4$ \citep{2002ApJ...568..324P}. 

Gravitational settling and element diffusion result in heavier elements sinking from the star's surface over long timescales.
Therefore, we turn on elemental diffusion for our models. Conversely, Procyon B is metal polluted; these effects alone cannot account for its atmospheric abundances.

\begin{figure*}
    \subfigure[]{\includegraphics[scale=0.58]{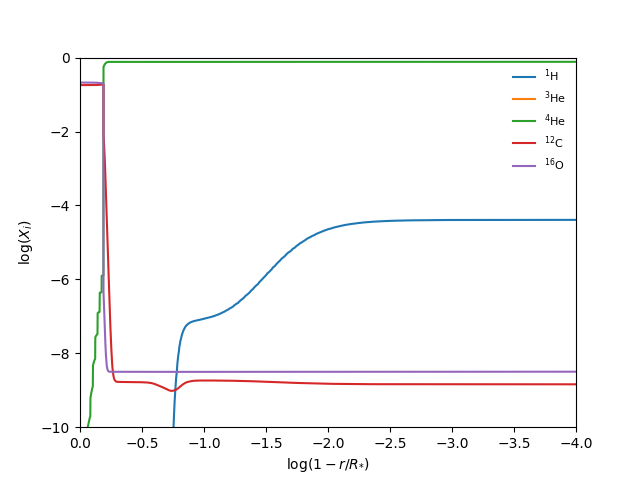}}
    \subfigure[]{\includegraphics[scale=0.35]{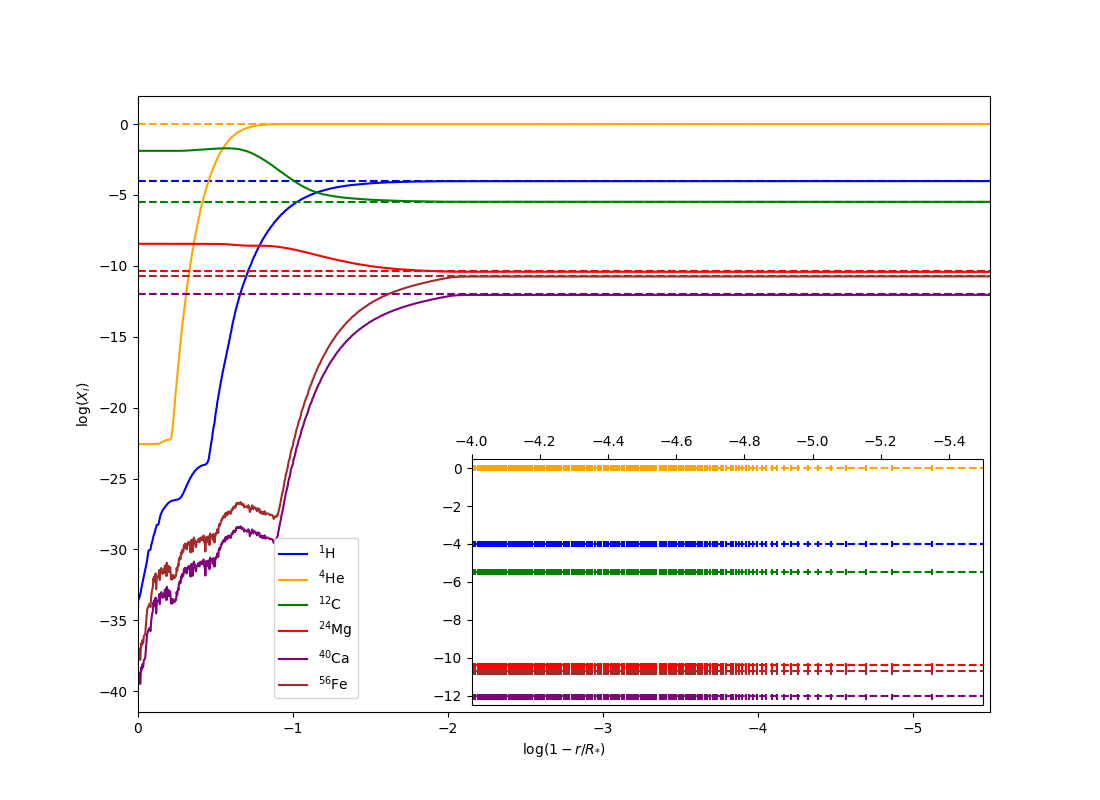}}

\caption{The interior chemical profile for the best fitting model of Procyon B is shown, with the $+x$ direction being towards the surface. We include the most relevant isotope abundances calculated by MESA (a) as well as the abundances of the accreted materials after the accretion episode during the WD cooling phase (b). These isotopes were accreted at a constant rate of $\sim1\times10^{-15}$\Msun/yr. Dashed lines on the right plot represent upper limits from \cite{2002ApJ...568..324P}.}
\label{chemical_profiles}

\end{figure*}

To fit our WD models to observed abundances we first strip the H envelope using MESA's \texttt{relax\_mass\_to\_remove\_H\_env} mass loss function. We then model the accretion of C and other metals directly onto the surface of the WD via MESA's composition controls. We started with initial mass fractions of $0.99$ for C, $\sim1.0\times10^{-4}$ for Fe and Mg, and $\sim1.0\times10^{-6}$ for Ca in an attempt to mimic current relative surface abundances. The rate of accretion was then adjusted until abundances of the final profile reached the upper limits from \cite{2002ApJ...568..324P}. We were able to recreate observed abundances at a constant accretion rate of $\sim1\times10^{-15}$\Msun/yr over the cooling lifetime of the white dwarf ($\sim1$Gyr). The exact source of this ``reservoir'' of metals is outside the scope of this paper however. Afterwards, we place the star back on the cooling sequence until it reaches the observed luminosity \citep{2012ApJS..199...29G}. We display the chemical profiles of the best model in Figure~\ref{chemical_profiles}.

\subsection{The Born Again Scenario}
The best theory to explain the formation of hydrogen-deficient white dwarfs is the born again scenario, through a very late thermal pulse (VLTP). In this model, the expansion of the helium burning layer intruding upon the hydrogen-rich envelope causes the hydrogen to be burned rapidly. This leads to the expansion of the proton-burning convection zone, and an increase in temperature, moving the star back to the asymptotic giant branch. Ultimately, this results a hydrogen deficient white dwarf, with photospheres consistent with the observations of PG~1159 stars \citep{bertolami_2006}. At the current \teff\ of Procyon B, diffusion and gravitational settling are dominant and any spectroscopic signature due to the VLTP would not be observed.  

The major challenge of computing a grid of models with the inclusion of the born again scenario is convergency, which requires a much smaller mesh-point, therefore much more computational time. Since the born again in reality is expected to last hundreds of years \citep{bertolami_2006}, the computation time is at least one order of magnitude longer than models where the outer layers are removed after the asymptotic giant branch phase. The scope of our study is to explore a multi-parameter space and their impact on stellar evolution constrained by observables. Moreover, through these models, we aim to constrain the initial mass of Procyon~B and the age of the system. We have computed a coarser grid with the inclusion of VLTPs (see section ~\ref{sec:bornagain} for details), centered around the best model without any VLTP. 

\section{Discussions}

\begin{figure*}[t!]
\hfill
\subfigure{\includegraphics[width=8cm]{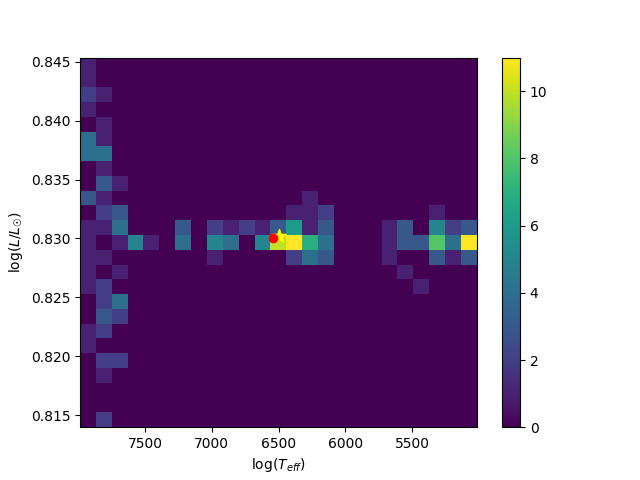}}
\hspace{1em}
\subfigure{\includegraphics[width=8cm]{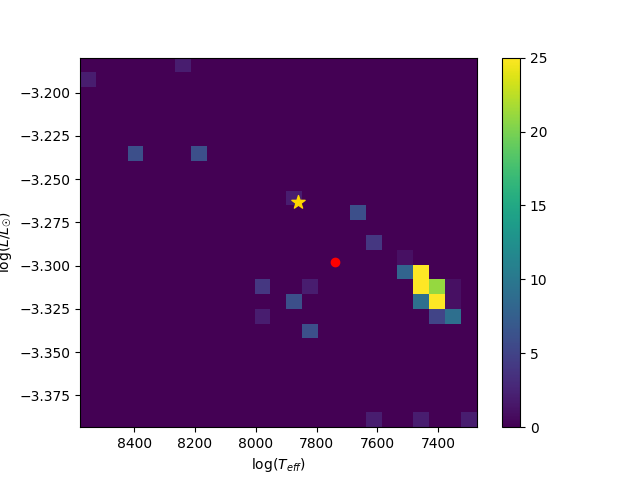}}
\hfill
\caption{2D histogram of the number of MESA models calculated for Procyon~A (left panel) and B (right panel) on the H-R diagram. From the color scale, the yellow bin represents the largest  number of models for a given combination of $L$ and \teff. The gold star indicates our best model and the red dot is the photospheric observations.}
\label{Procyon_2d_hist}
\end{figure*}

From the observational constraints in previous studies, we created $\sim500$ MESA models for the binaries in the Procyon system. The parameters of the grid are summarized in Table~\ref{parameters}, including the boundaries and the steps for each physical quantity. We display 2D histograms of the number of MESA models calculated for Procyon~A (left panel) and B (right panel) on the H-R diagram in Figure~\ref{Procyon_2d_hist}. From the color scale, the yellow bin represents the largest number of models for a given combination of $L$ and \teff. 

\begin{table}[]
\hspace{-1.5cm}
    \centering
    \begin{tabular}{||c|c|c||}
          \hline
 Parameter&Range& Step Size \\ [0.5ex] 
 \hline\hline
 $M_{Ai} (M_{\odot})$& 1.411 - 1.531 & 0.02 \\ 
 \hline
$M_{Bi} (M_{\odot})$& 1.9 - 2.6 & 0.1  \\
 \hline 
 $Z_{i}$ & 0.0125 - 0.0275 & 0.003 \\
 \hline
 $\alpha$ & 1.7 - 2.5 & 0.4\\
 \hline
 $\beta$ & 0.0 - 1.0 & 0.5\\
 \hline
    \end{tabular}
    \caption{Parameter space for MESA models calculated for Procyon system, constrained by previous observations.}
    \label{parameters}
\end{table}

In this section, we discuss how initial mass, metallicity, core overshoot, and mixing length impact stellar evolutionary models and the derived quantities.

\subsection{Age dependency on convection}

\begin{figure*}[ht!]
\centering
\hspace{-2cm}
\includegraphics[scale=0.7]{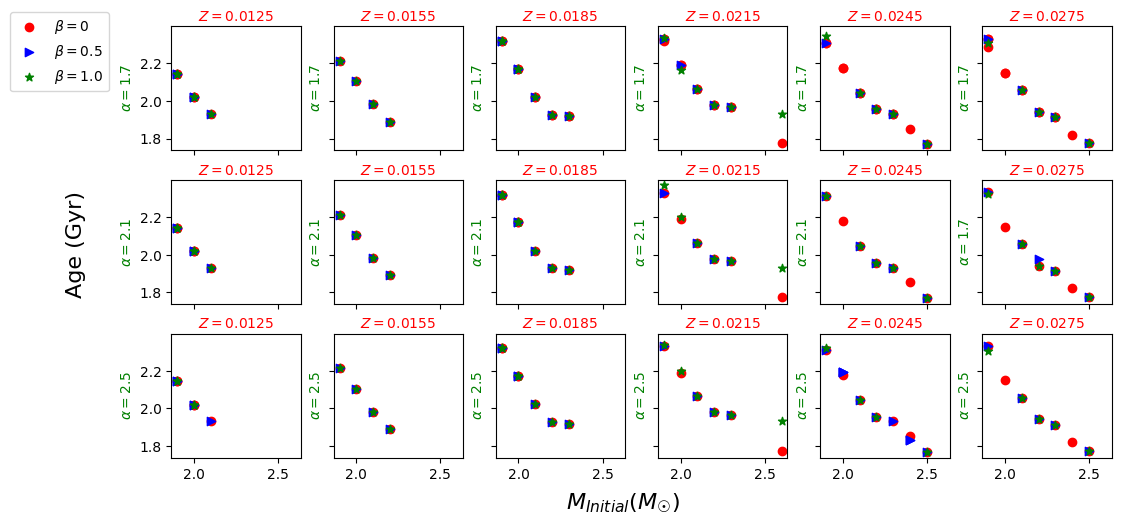}
\caption{Results from our MESA models for the age of the white dwarf (in Gyr) as a function of intial mass (in \Msun). Each symbol represents a different value of core overshoot $\beta$, where the red circles are $\beta=0$, the blue triangles $\beta=0.5$, and the green stars $\beta=1.0$.
Each row are models for a fixed $\alpha$, from 1.7 in the top panel, 2.1 in the middle, and 2.5 in the bottom. Each column displays models for fixed metallicity with $Z=0.0125$ in the left panels increasing when moving rightwards to 0.0275. Models that did not converge or with ages below $\sim1.8$ Gyr are not shown.
As expected, the age of Procyon B decreases with the increase of initial mass. For a given metallicity, the age is much less sensitive to the values of $\alpha$ and $\beta$ than it is to the initial mass.}
\label{agevsmassbeta}
\end{figure*}

The amount of time stars spend in each evolutionary phase is mainly determined by their initial masses, which in turn determines the amount of available material as stellar fuel. Convection, namely mixing length free parameter $\alpha$ and core overshoot $\beta$, plays a secondary role in the reserves of energy of stars, which impacts directly the timescales of each evolutionary phase. In Figure~\ref{agevsmassbeta}, we show the comparison of age for different $\beta$ values across all computed models for Procyon~B, for each combination of metallicity $Z$ and $\alpha$.
 
\subsection{IFMR of Procyon~B}\label{sec:ifmr}

\begin{figure*}[ht!]
\centering
\hspace{-4cm}
\includegraphics[scale=0.75]{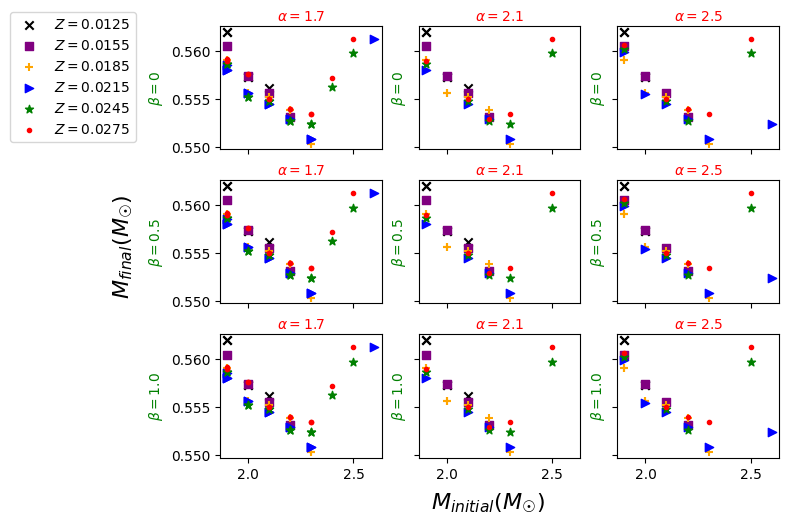}
\caption{Results from the model grid showing how the white dwarf IFMR changes depending on the choice of $Z$, $\alpha$, and $\beta$. Most models lie within the region of $\sim 0.55 - 0.57$ \Msun, with those with lower mixing able to produce more higher mass models at large metallicities, likely as a result of more models converging to a solution.}
\label{procyon_ifmr_split_z}
\end{figure*}

The IFMR quantifies the total amount of mass loss a star will undergo through its evolution. 
From our computed model grid, we plot the progenitor mass as a function the final white dwarf mass to create our own theoretical IFMR between $M=1.9-2.6$\,\Msun\ for Procyon~B displayed in Figure~\ref{procyon_ifmr_split_z}. We separate the models by combinations of $\alpha$ values (columns) and $\beta$ values (rows) for different metallicities (different symbols) to analyze the IFMR dependency on these parameters.  
While the overall shape of the IFMR does not change notably, we note a scatter of $\sim0.001$\,\Msun\ in $M_\mathrm{final}$. It is important to mention that not all combinations of parameters produced converging models for a given initial mass. 

Lastly, we notice that the IFMR is not the monotonically increasing \citep[e.g.][]{2018ApJ...866...21C}, but shows a slight decrease, consistent to the kink observed at $M\simeq 2$\,\Msun \citep{2020NatAs...4.1102M}. The reason for this arises from a complex interplay of several physical parameters and processes during the thermally pulsing asymptotic giant branch phase. \cite{mariago_2022} determined that the most important dependencies and parameters are the efficiency of the third dredge-up and the mass loss rates for carbon stars. This carbon enrichment affects the composition of the star, influencing opacity, and thus \teff, its dust-driven mass loss rate, and the core mass. As a result of this dredge-up mechanism, the core mass, and by proxy the total mass, gets smaller as our initial mass goes up. 

\section{Determination of the Best Model}\label{sec:results}

To determine the best evolutionary model that fits the observations, we implemented a minimization, based on equation \ref{chi}, where $\xi$ represents the following physical quantities: stellar mass ($M$), surface luminosities ($L$), and  effective temperatures (\teff ). The observed values are listed in Section~\ref{observ} and the model refers the our computed MESA model grid. In this formulation, the minimization of $\chi$ represents the best model.  

\begin{equation}
\chi = \sqrt{\sum_i\left(\frac{\xi_{i,\mathrm{observed}}-\xi_{i,\mathrm{model}}}{\xi_{i,\mathrm{observed}}}\right)^2}
\label{chi}
\end{equation}

\begin{figure*}[ht!]

\subfigure{\includegraphics[width=8cm]{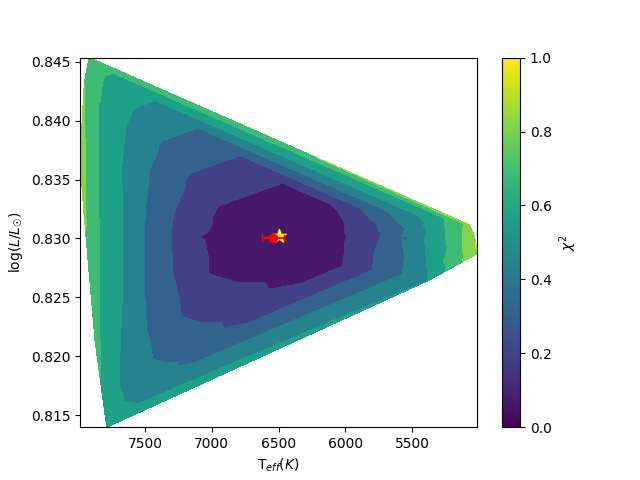}}
\hfill
\subfigure{\includegraphics[width=8cm]{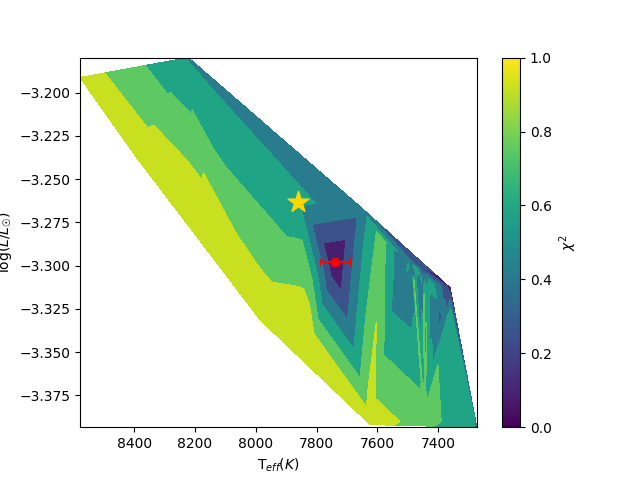}}
\hfill
\subfigure{\includegraphics[width=8cm]{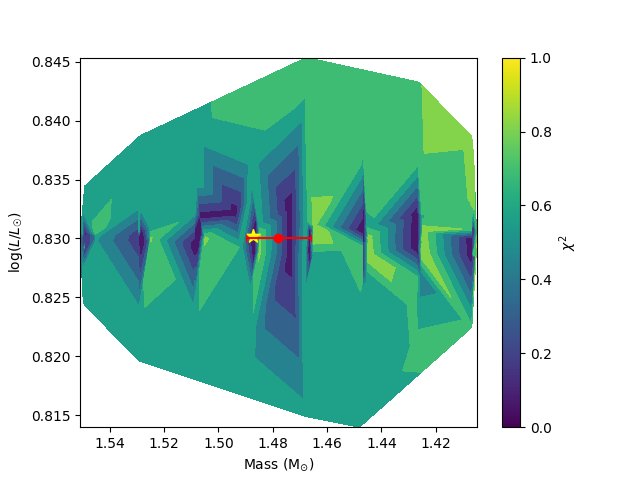}}
\hfill
\subfigure{\includegraphics[width=8cm]{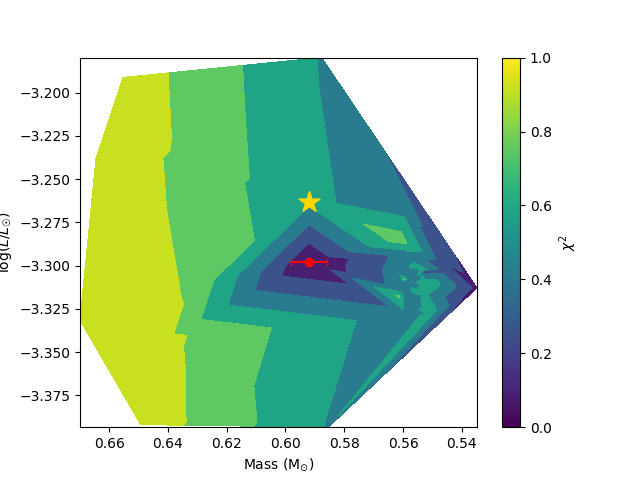}}
\hfill
\subfigure{\includegraphics[width=8cm]{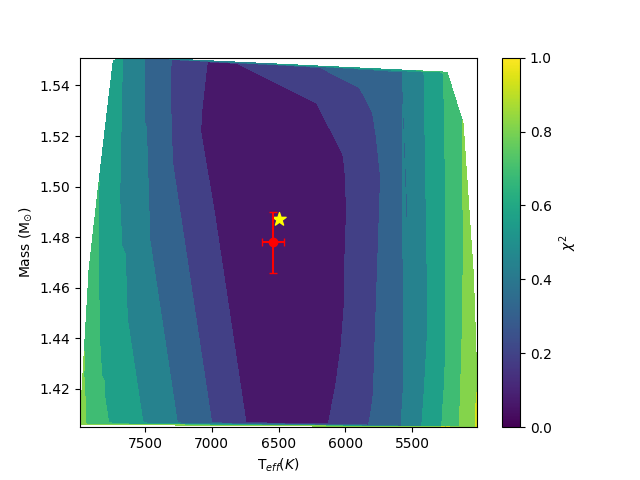}}
\hfill
\subfigure{\includegraphics[width=8cm]{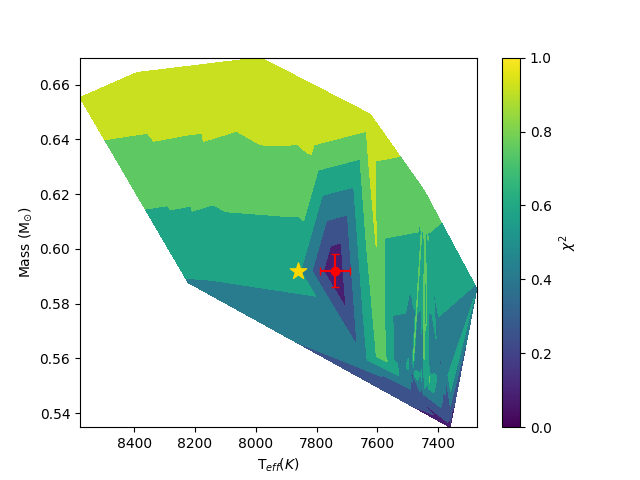}}
\caption{Results of the comparison of the observations to the models for Procyon~A and B, in the parameter space of $M$, \teff, and \loglum. The plots in the left columns are for Procyon~A, while the right ones for B. The color scale from dark blue to yellow represents the calculated $\chi^2$ for each model from lowest to highest. The gold star is the model with the smallest $\chi^2$ value, and therefore, our best fit to the observations (red dot).}
\label{contour_plots}
\end{figure*}

The results are plotted in Figure~\ref{contour_plots}, showing how the $\chi^2$ values vary as a function of $M$, \loglum, and \teff. The best fits are models represented by the darker blue. 

\begin{figure*}[t!]
\centering
\includegraphics[scale = 0.5]{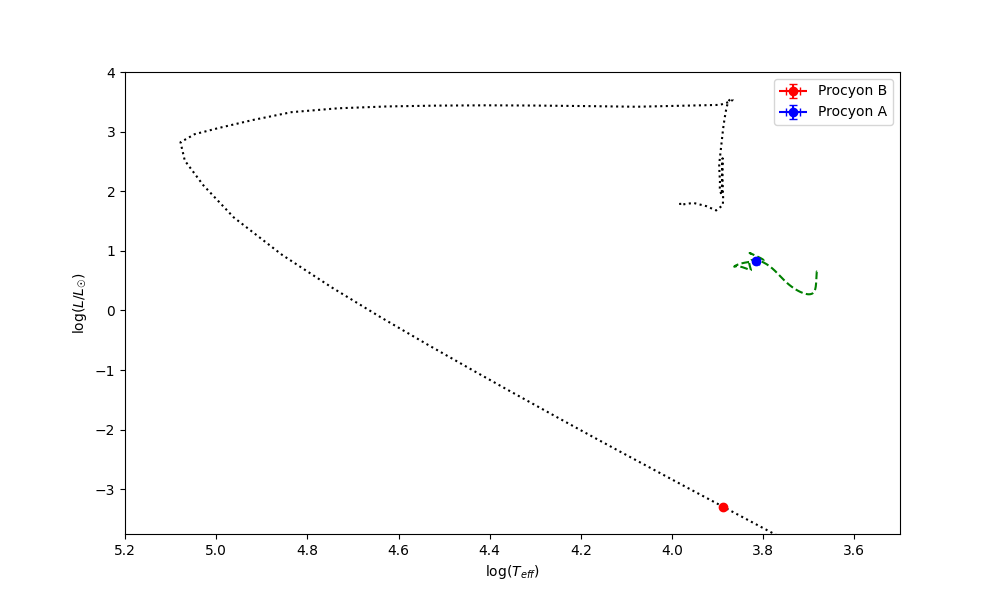}
\caption{H-R Diagram for the best model of the Procyon system. The evolutionary track of Procyon~A is shown in green, long-dashed lines, while the track for B is in black, dotted lines. The position of both stars, A in blue and B in red, are within the observed uncertainties for luminosity and temperature.}
\label{hr_best}
\end{figure*}

For Procyon~A, our best model corresponds to $M_\mathrm{A}=1.487\pm0.103$\,\Msun, \teffa$=6943 \pm40$\,K, \logluma$=0.830 \pm 0.005$, log$(R/R_{\odot})=0.721\pm0.003$, with convective parameters $\alpha = 2.5$ and $\beta = 0.5$. 
Our best model for Procyon~B has a progenitor mass of $2.2\pm0.3$\,\Msun\ and a final white dwarf mass of $0.592\pm0.082$\,\Msun, \teffb = $7861\pm251$\,K,  \loglumb =$-3.263\pm0.179$, log$(R_\mathrm{B}/R_{\odot})=-1.900\pm0.035$, and convection parameters $\alpha = 2.5$, and $\beta = 1.0$. The best fit metallicity for both Procyon~A and B is $Z= 0.0245\pm0.0077$. Our best models result in a total age of the system of $2.23\pm0.90$ Gyr, with a cooling age of $1.197\pm0.495$ Gyr for Procyon~B.

The uncertainties for all physical quantities were calculated by comparing the difference of the best model to the next best model $(d-d_0)$ and their least squares $\chi_0$ and $\chi$, following equation \ref{error_eq} \citep{1986ApJ...305..740Z}

\begin{equation}
    \label{error_eq}
    \sigma=\sqrt{\frac{(d-d_0)^2}{\chi-\chi_0}}
\end{equation}


\subsection{Models with very late thermal pulse}\label{sec:bornagain}

\begin{figure*}[t!]
\centering
\includegraphics[scale = 0.5]{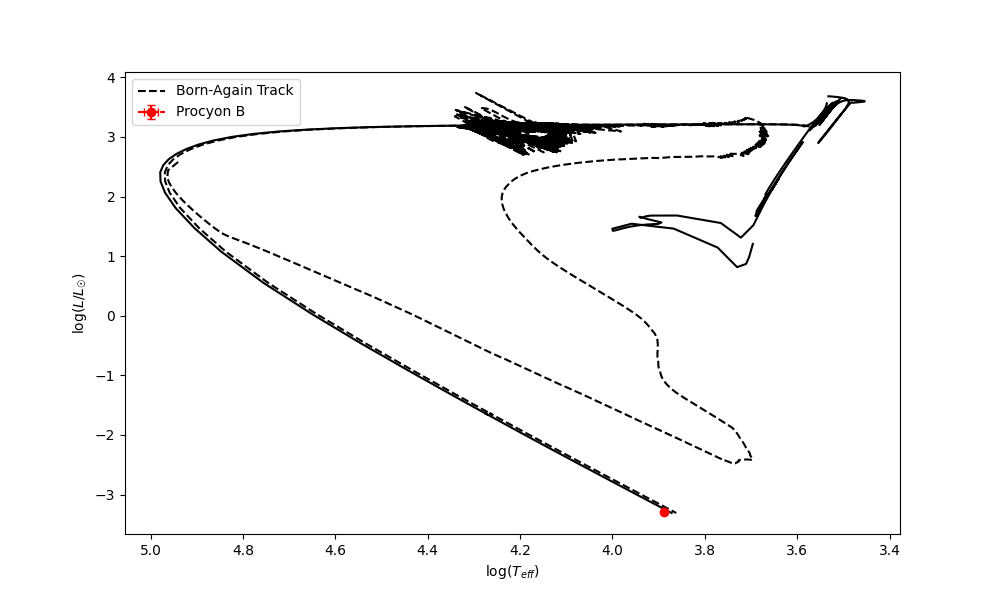}
\caption{Best model from the coarse grid including a very late thermal pulse, which creates the white dwarf through the born again scenario (dashed line), in comparison to the best model without this prescription (solid line) and the current position of Procyon~B in the H-R diagram.}
\label{hr_best_born_again}
\end{figure*}

We create a smaller, coarser grid of models where hydrogen-deficient white dwarfs are formed through the born again scenario. We chose an initial mass range centered on the best solution for progenitor mass $M_\mathrm{B} = 2.2 $ \Msun, with $\Delta M = 0.1$ \Msun\, spanning the range $1.9 - 2.5$ \Msun. We hold the other properties ($Z,\alpha,\beta$) constant, with values of 0.0245, 2.5, and 1.0 respectively.
In this subset of models, we initialize a VLTP after the asymptotic giant branch phase. We relax the mass via a wind mass loss function until the star becomes a white dwarf.
For opacities, we use the Type 2 OPAL tables from \cite{1996ApJ...464..943I} which include enrichment from C and O. For low temperatures ($\log T_\mathrm{eff} \lesssim 4$) we use opacities from \cite{ferguson_2005}.
Following the methods of \cite{2025ApJ...995..114L}, we remove the envelope during the thermal pulse cycle at $\sim \log\left(\frac{L_{\mathrm{He}}}{L_{\odot}}\right)=2.3$. We then compute models down to the observed luminosity and, once again fit, to the observations via $\chi^2$ minimization. Our best model is presented in Figure~\ref{hr_best_born_again}. This model has an identical mass to our previous best solution with $M_\mathrm{initial} = 2.2$ \Msun\, and $M_\mathrm{B} = 0.59$ \Msun. The radius is slightly larger however, with log$(R_\mathrm{B}/R_{\odot})=-1.89$ and a slightly lower \teff$=7601$\,K, meaning that both these quantities differ by less than 5\% in the best model without born again scenario. The quantity with the largest difference ($\sim 7$\%) is age, as the best model of this grid results an age of 2.4\,Gyr. In Table~\ref{table_born_again}, we outline the best model fits for Procyon B, using evolutionary models with and without the born again prescription, which fits better the observations (smaller $\chi$ value).

\begin{table*}[ht]
\centering
\hspace{-2cm}
\begin{tabular}{llllllll}
\toprule 
&&\multicolumn{3}{l}{Extensive model grid} & \multicolumn{3}{c}{Models with born again }\\
\toprule
$M_\mathrm{initial}$ (\Msun) & $M_\mathrm{final}$ (\Msun) & \teff (K)& Age (Gyr) & $\chi$ & \teff (K) & Age (Gyr) & $\chi$ \\
\hline
1.9 &0.560&7465&2.32&$0.069$&7629&2.65&$0.098$ \\  
2.0 &0.578&7369&2.20&$0.105$&7630&2.43&$0.143$ \\
2.1 &0.547&7437&2.04&$0.081$&7611&2.31&$0.133$ \\
2.2 &0.592&7861&2.23&$0.019$&7601&2.40&$0.042$ \\
2.3 &0.524&7429&1.93&$0.085$&7599&2.16&$0.153$ \\
2.4 &0.575&7482&1.83&$0.069$&7613&2.06&$0.089$ \\
2.5 &0.597&7481&1.77&$0.069$&7624&2.04&$0.080$ \\

\toprule

\end{tabular}
\caption{Best models for different $M_\mathrm{initial}$ for Procyon B, from the comparison of the observations to the extensive model grid and the subset of models computed with the very late thermal pulse, forming the white dwarf in the born again scenario. These models were computed using the best values from the extensive model grid, as discussed in Section~\ref{sec:results}: $Z = 0.0245$, $\alpha = 2.5$, and $\beta = 1.0$. Independent of the evolutionary prescription used to form a hydrogen-deficient white dwarf, the best models are for the exact same mass: $M_\mathrm{initial}=2.2$\,\Msun\ and $M_\mathrm{final}=0.592$\,\Msun. The differences in \teff\ are $\sim5\%$ while most ages are around $\sim10\%$ different.}
\label{table_born_again}
\end{table*}

\section{Comparisons to independent determinations of the evolution of the Procyon system}

The derived quantities from fitting evolutionary models using observational constraints are age of the system and progenitor mass for Procyon~B. In this section, we compare our modeling of Procyon system to previous studies. 

\subsection{Comparisons of System Ages}

Age is a fundamental parameter in stellar evolution because it constrains the timeline for all physical processes in a star's life. Despite its importance, age is a derived, model dependent quantity. It can be obtained from evolutionary tracks, depending on the observed photospheric parameters, and through asteroseismology and gyrochronology. 
In our analysis, the best model for Procyon yields the age of the system to be $2.23\pm0.90$\,Gyr, where the time on the main sequence of Procyon~B is $1.043\pm0.495$\,Gyr, corresponding with a cooling age of $1.197\pm0.495$\,Gyr.

There is no reliable gyrochronology determination of age for Procyon~A because it is a relatively old star, and because its rotation has slowed to the point where the dynamo that drives the magnetic braking process is significantly disrupted. This makes its rotation period a poor indicator of its age. Instead, its age is determined from comparisons with isochrones and evolutionary track models, which are constrained by asteroseismology. \cite{Guenther_2014} modeled Procyon~A with a Bayesian approach and calculated evolutionary tracks using YREC \citep{2008Ap&SS.316...31D}, varying $M$, $Y$, $Z$, $\alpha$, and $\beta$, deriving from their most probable model an age within the range 2.4 -- 2.8\,Gyr, for $1<\beta<1.5$. Increasing the amount of overshoot does affect the mass of the convective core and the age. For instance, their best model without overshoot, $\beta=0$, yields an age of $\sim1.85$\,Gyr.   
A somewhat independent determination of the age of Procyon~A is $\sim2.7$\,Gyr \citep{2015ApJ...813..106B}, from the comparison of similar evolutionary tracks with large overshooting \citep{Guenther_2014}. Our age is slightly smaller than these values, but consistent with the fact that our best model contains high overshoot. 

An alternative way to estimate the age of the system is fitting the cooling age of Procyon~B. \cite{2015ApJ...813..106B} used their dynamical mass of $0.592\pm0.006$\,\Msun\ and photospheric parameters to compare with the white dwarf cooling tracks \citep{1995PASP..107.1047B}, deriving $1.37\pm0.04$\,Gyr as its cooling age. Their determination is larger than ours, which is an consistent with their overall age of the system being larger as well. 

Lastly, we must discuss the timescales involved with the born-again scenario. The time between the start of the He flash on the white dwarf cooling track, to the ignition of the H-shell, to the onset of pulsations and beyond, is completely negligible in terms of the lifetime of the star. However, our models with the born-again scenario are $\sim 0.20$ Gyr older than their equivalents without. This is due to post-AGB late thermal pulses. The models with these extra LTPs and VLTPs have a system age closer to the determinations from \cite{2015ApJ...813..106B} and the high overshoot grids from \cite{Guenther_2014}, supporting the idea of VLTPs in Procyon's evolutionary history.

\subsection{Comparisons of Progenitor Masses}

The study by \cite{Liebert_2013} indicate a shorter cooling age of $1.19\pm0.11$\,Gyr for Procyon~B, which implies a progenitor mass of $2.59^{+0.44}_{-0.26}$\,\Msun. Their result is larger than our determination of $2.2\pm0.3$\,\Msun. However it is important to point out that their cooling age is shorter, which comes from their model assumption of no core overshoot.  \cite{2015ApJ...813..106B} determines a range of progenitor masses from $1.9-2.20$\,\Msun, depending on the amount of overshoot in the models. Our best model for Procyon~B has convective parameters $\alpha=2.5$ and $\beta=1.0$, similar to their models. Our results are consistent with previous determinations, which validates our approach in modeling Procyon's evolution. 

\section{Conclusion}
In this work, we presented our description of an extensive evolutionary MESA model grid calculated for Procyon~A and B, constrained by precise dynamical masses and spectroscopic parameters. Our primary goal was to quantify how key physical parameters--metallicity ($Z$), mixing length ($\alpha$), and core overshoot ($\beta$)--affect the inferred system age and the progenitor mass of Procyon~B, and to establish Procyon as a benchmark for future studies of hydrogen-deficient white dwarfs.
We have used the published values for the observational parameters and some derived quantities to constrain the boundaries of our grid specifically for this system, similar to the approach of \cite{2024ApJ...977...41K} for Sirius. We modeled this system as non-interacting due to the known orbital solution of Procyon. Our best fits yielded a mass of $1.487 \pm 0.095$\,\Msun\ for Procyon~A and $0.592 \pm 0.082$\,\Msun\ for Procyon~B.

Within the explicitly stated scope of our models, which do not include a very-late thermal pulse (“born-again”) evolution, we find that:
\begin{itemize}
\item The best-fit solution reproduces the observed properties of both components, yielding $M_A = 1.49\,M_\odot$, $M_B = 0.592\,M_\odot$, a total age of $2.23$~Gyr, and a white dwarf cooling age of $1.20$~Gyr, consistent with independent determinations.
\item Sensitivity analysis shows that convective overshoot and metallicity dominate the uncertainty budget for age and progenitor mass, underscoring the need for additional constraints such as asteroseismology for Procyon~A.
\item Our preliminary initial-to-final mass relation (IFMR) for helium-dominated white dwarfs in the $1.9$–$2.6\,M_\odot$ progenitor range exhibits mild non-monotonicity near $M_{\rm init}\!\sim\!2\,M_\odot$, qualitatively consistent with predictions from third dredge-up and carbon-star mass-loss physics.
\end{itemize}

For the models that did include the born again scenario and VLTPs, we found that they were $\sim0.2$ Gyr older than their counterparts without. Aside from this, and their evolution on the H-R diagram, there were no large differences in final physical parameters nor the $\chi^2$ between the two sets of models. Post-AGB LTP however, do lead to longer model lifetimes, supporting the findings of other studies that report system ages longer than $\sim2$ Gyr. 

We also studied the sensitivity of our models to multiple parameters that influence stellar evolution. Namely, we investigated how the convective parameters, $\alpha$ and $\beta$, as well as metallicity $Z$ affect the age and the progenitor mass of Procyon~B. 

With our analysis, we were able to constrain the age of the system of $2.23\pm0.90$\,Gyr and a cooling age of Procyon~B of $1.197\pm0.495$\,Gyr. Comparing our determinations with previous studies, we validated our analysis, as our ages are consistent with independent modeling. We also discussed that core overshoot is an important parameter in affecting the age, which can easily explain discrepancies with these values in the literature. 

Finally, our best model fit yield a progenitor mass of $2.2\pm0.3$\,\Msun\ for Procyon~B. Our result places important constraints on the IFMR for hydrogen-deficient white dwarfs. There are very few determinations for this class of white dwarfs, especially when compared to the most numerous spectral types with hydrogen-dominated photospheres. More importantly, our results shed light on the evolutionary differences between white dwarfs with different photospheric compositions. 


If more precise observables become available in the future, we can expand our model grid, constrain the parameters more tightly, and possibly vary other quantities to better model Procyon and other non-interacting ``Sirius-like'' systems.

\bibliography{procyon_system}
\bibliographystyle{aasjournal}

\end{document}